\documentclass[%
twocolumn,
 reprint,
 amsmath,amssymb,
 aps,
]{revtex4-1}

\usepackage{graphicx}
\usepackage{dcolumn}
\usepackage{bm}
\usepackage{hyperref}

\hypersetup{
	colorlinks=true,
	linkcolor=blue,
	filecolor=magenta,      
	urlcolor=blue,
	citecolor=blue,
	anchorcolor=blue,
}

\begin{document}

\noindent 
{\bf Reply to ``Comment on `Anomalous Reentrant 5/2 Quantum Hall Phase at Moderate Landau-Level-Mixing Strength' "}\\


In Ref.~\cite{Simon} (Comment) Simon has raised an objection that the proposed wavefunction [Eq.~2] (WF-I) in Ref.~\cite{DDM} which is antisymmetrization of so-called Halperin 113 wave function (WF-II) \cite {Halperin} does not describe a fractional quantum Hall (FQH) state.  
Simon conjectures that antisymmetrization will have a minor effect on a phase separated state. However, the indistinguishability of particles cannot be treated naively as antisymmetrization often makes a drastic change in the characteristic of a quantum state. For example, antisymmetrization of Abelian Halperin 331 wavefunction \cite {Halperin} transforms it into a wavefunction for non-Abelian state \cite{Cappelli}.

\begin{figure}[h]
	\centering
	\includegraphics[width=\linewidth]{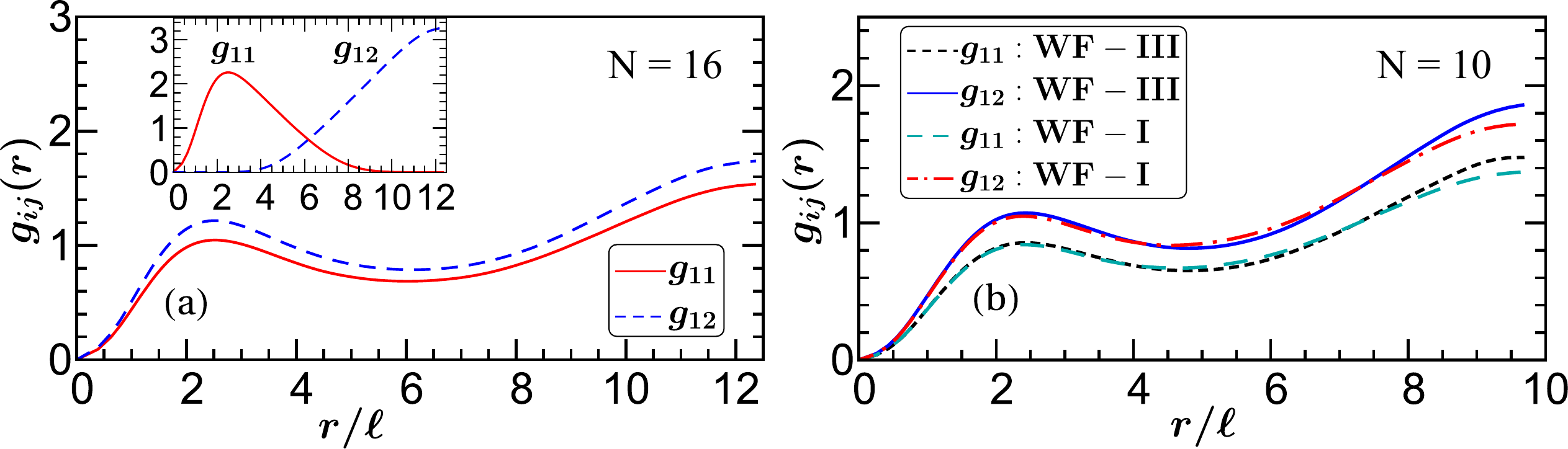}
	\caption{(a) $g_{11}$ and $g_{12}$ for WF-I. Inset: Same for WF-II. (b) $g_{11}$ and $g_{12}$ for WF-I and WF-III are compared.}
	\label{fig.pairgrid}
\end{figure}

It is natural that two species in WF-II will avoid each other as they feel more repulsion than the particles between same species.
In WF-I, however, being all the particles indistinguishable in nature, {\em any} $N/2$ particles in an $N$ particle system can form a group.
Therefore, formation of two distinct physically separated groups is not obvious. Instead, all the particles are expected to optimize their positions for forming a uniform liquid as in any of the known FQH states. 
To this end, let us consider $1$ through $N/2$ particles in the first group and the remaining particles in the second group. Both homogeneous, $g_{11}$, and heterogeneous, $g_{12}$, pair correlation functions [Fig.~\ref{fig.pairgrid}(a)] hover around $1.0$, as in any other FQH liquid.
In contrast, as shown in Ref.~\cite{Goerbig}, the behaviors of $g_{11}$ and $g_{12}$ indicate phase separation [inset of Fig.~\ref{fig.pairgrid}(a)] between two groups or species.

The entanglement spectra (ES) for a toy-model numerical wavefunction (WF-III) is shown \cite{Simon} as identical to the same for the $\mathcal{A}$ phase in Ref.~\cite{DDM} for anti-Pfaffian flux, $N_\Phi = 2N+1$. 
However, for the particle-hole symmetric 
Pfaffian (PH-PF) 
flux $N_\Phi = 2N-1$ (same for the WF-I), the ES does not match, because the ${\cal A}$ phase has a high degree of particle-hole asymmetry.
Nonetheless, WF-III reproduces almost identical $g_{11}$ and $g_{12}$ to the same for WF-I [see Fig.~\ref{fig.pairgrid}(b)]. 
Therefore, WF-III also describes a liquid state, and may also be topologically identical to the ground state wavefunction of the $\mathcal{A}$ phase. 
It is, however, not surprising that an $L=0$ wave function which is {\em not} the global ground state of a toy-model Hamiltonian can be representative of the {\em global} ground state of a realistic Hamiltonian, because such a wavefunction does not necessarily reflect the nature of the toy model considered.

\begin{figure}[h]
	\centering
	\includegraphics[width=\linewidth]{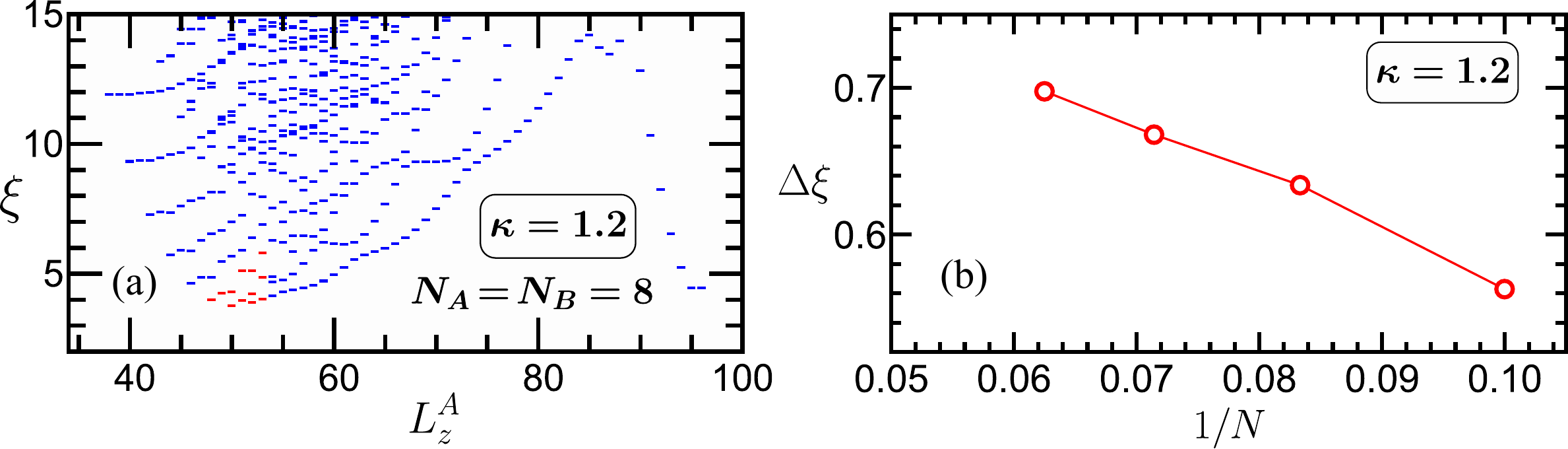}
	\caption{(a) ES %
		in the $\mathcal{A}$ phase at PH-PF flux. (b) Scaling of $\Delta \xi$
		at PH-PF flux shift with $1/N$ in the $\mathcal{A}$ phase.}
	\label{fig.esn16scaling}
\end{figure}

As far as the ES with an equal number of particles in two parts is concerned,  the Hilbert space of the $\mathcal{A}$ phase or WF-I allows all the particles to occupy up to the highest possible $N/2$ orbitals. We thus have the ES up to the highest possible $L_z^A$. While the number of states at higher possible $L_z^A$ is always finite in number, the number of states grows exponentially with $N$ for the medium range of $L_z^A$ for which the particles mostly occupy mid-level orbitals in the sphere. Therefore, the latter kind of states are more entangled and hence lower their lowest energies with the increase of $N$. On the other hand, the former kind of states, being least entangled, either increase their entangled energies or their separation with respect to the lowest entangled energy grows with $N$. In view of this,  we show the ES 
for $N=16$ and $N_\Phi =31$ 
in Fig.~\ref{fig.esn16scaling}(a) where the low energy branch at higher $L_z^A$ is much reduced, while universal low-lying edge state counting 1-1-2-2... remains unaltered. Further, the entangled energy difference $(\Delta \xi)$ between the lowest energy and the energy at the highest $L_z^A$ increases with $N$ [see Fig.~\ref{fig.esn16scaling}(b)].

To summarize, the proposed wavefunction and also the $\mathcal{A}$ phase in Ref.~\cite{DDM} describe the FQH liquid state rather than a phase separated  or stripe or bubble state. \\

\noindent Sudipto Das, Sahana Das, and Sudhansu S. Mandal\\
 \hspace{1cm} Department of Physics, Indian Institute of Technology Kharagpur, West Bengal 721302, India

\end{document}